# 33% Giant Anomalous Hall Current Driven by both Intrinsic and Extrinsic Contributions in Magnetic Weyl Semimetal Co$_3$Sn$_2$S$_2$

*Jianlei Shen, Qingqi Zeng, Shen Zhang, Hongyi Sun, Qiushi Yao, Xuekui Xi, Wenhong Wang, Guangheng Wu, Baogen Shen, Qihang Liu, Enke Liu*[*]

Jianlei Shen, Qingqi Zeng, Shen Zhang, Xuekui Xi, Wenhong Wang, Guangheng Wu, Baogen Shen, Enke Liu
State Key Laboratory for Magnetism, Institute of Physics, Chinese Academy of Sciences, Beijing 100190, China
E-mail: ekliu@iphy.ac.cn

Jianlei Shen, Qingqi Zeng, Shen Zhang
University of Chinese Academy of Sciences, Beijing 100049, China

Enke Liu, Wenhong Wang
Songshan Lake Materials Laboratory, Dongguan, Guangdong 523808, China

Hongyi Sun, Qiushi Yao, Qihang Liu
Shenzhen Institute for Quantum Science and Technology and Department of Physics, Southern University of Science and Technology (SUSTech), Shenzhen 518055, China

Qihang Liu
Guangdong Provincial Key Laboratory for Computational Science and Material Design, Southern University of Science and Technology, Shenzhen 518055, China

**Abstract**

Anomalous Hall effect (AHE) can be induced by intrinsic mechanism due to the band Berry phase and extrinsic one arising from the impurity scattering. The recently discovered magnetic Weyl semimetal $Co_3Sn_2S_2$ exhibits a large intrinsic anomalous Hall conductivity (AHC) and a giant anomalous Hall angle (AHA). The predicted energy dependence of the AHC in this material exhibits a plateau at 1000 $\Omega^{-1}$ cm$^{-1}$ and an energy width of 100 meV just below $E_F$, thereby implying that the large intrinsic AHC will not significantly change against small-scale energy disturbances such as slight $p$-doping. Here, we successfully trigger the extrinsic contribution from alien-atom scattering in addition to the intrinsic one of the pristine material by introducing a small amount of Fe dopant to substitute Co in $Co_3Sn_2S_2$. Our experimental results show that the AHC and AHA can be prominently enhanced up to 1850 $\Omega^{-1}$ cm$^{-1}$ and 33%, respectively, owing to the synergistic contributions from the intrinsic and extrinsic mechanisms as distinguished by the TYJ model. In particular, the tuned AHA holds a record value in low fields among known magnetic materials. This study opens up a pathway to engineer giant AHE in magnetic Weyl semimetals, thereby potentially advancing the topological spintronics/Weyltronics.

# 1. Introduction

As an important electronic phenomenon, the anomalous Hall effect (AHE) has attracted extensive research attention from the perspectives of fundamental physics and technical applicability.[1-3] Large AHE are essential for applications such AHE sensors[4] and those based on spin-transfer torque.[5] As per current knowledge, it is believed that AHE originates from two different microscopic mechanisms, i.e., intrinsic and extrinsic ones. As regards the intrinsic mechanism, Karplus and Luttinger first proposed that the spin–orbit interaction together with interband mixing results in an anomalous electron velocity in the direction transverse to the electric field;[6] this mechanism was recently revived based on the framework of Berry curvature,[6-9] which yields the resistivity relation $\rho_H \propto \rho_{xx}^2$. As regards the extrinsic mechanism, Smit and Berger proposed skew scattering and side-jump effects in the scattering process that is affected by the spin–orbit interaction, which yield the relations $\rho_H \propto \rho_{xx}$ and $\rho_H \propto \rho_{xx}^2$, respectively.[10,11] However, a large number of theoretical calculations and experimental measurements indicate that the intrinsic contribution is the dominant cause for large AHE in most materials.[12,13] Therefore, a large AHE can be expected in materials with strong Berry curvature. In particular, in magnetic Weyl materials, strong Berry curvature can be induced by Weyl nodes and gapped nodal lines engendered by the spin–orbit interaction.[14-16]

Recently, topologically enhanced Berry curvature has been exploited to achieve intrinsic giant anomalous Hall conductivity (AHC, ~1130 $\Omega^{-1}$ cm$^{-1}$) and anomalous Hall angle (AHA, ~20%) in the magnetic Weyl semimetal Co$_3$Sn$_2$S$_2$,[17] in which the Weyl nodes, located only 60 meV above the Fermi level ($E_F$), act as magnetic monopoles and sources of Berry curvature, a form of pseudo-magnetic field in momentum space. Moreover, both angle-resolved photoemission spectroscopy (ARPES) and scanning tunneling spectroscopy (STM) have been used to confirm the existence of the Weyl phase.[18,19] The AHC measured in experiments is consistent with that of theoretical calculations from the band-structure Berry curvature, which indicates that the AHC mostly originates from the intrinsic component in Co$_3$Sn$_2$S$_2$.

However, the extrinsic contribution to AHE can be remarkable in certain systems. For example, in the topological ferromagnet Fe$_3$Sn$_2$, the extrinsic contribution can be five times larger than the intrinsic one.[20] In addition, a transition from intrinsic AHE to the extrinsic one has been reported in the magnetic Weyl semimetal candidate PrAlGe$_{1-x}$Si$_x$ for $x > 0.5$.[21] Furthermore, the magnetic Weyl system Co$_2$MnGa also shows a large AHC (1530 $\Omega^{-1}$ cm$^{-1}$), in which the extrinsic component accounts for nearly 50% of the AHC.[22] These results indicate that the extrinsic mechanism can also significantly contribute to the total AHE. Nevertheless, these reported systems still exhibit a relatively low AHA of <12%. Microscopically, the extrinsic mechanism is related to disorder scattering from the spin–orbit interaction of conduction electrons and impurities, and it strongly depends on factors such as crystal defects, impurity concentration, and film thickness.[23-27] Thus, the desired AHC and AHA can be enhanced by extrinsic contributions from suitable impurities.

# 2. Design scheme

To obtain enhanced AHE, in this study, we propose a scheme to induce the extrinsic component based on the intrinsic one in the magnetic Weyl semimetal $Co_3Sn_2S_2$. In this system,[17] the strong Berry curvature is dominated by the relatively clean topological bands around the Fermi level, including the Weyl nodes and gapped nodal lines from cobalt (Co) orbits, which produces a significant transverse Hall transport effect on the Co kagome lattices. As per our previous work,[17] the energy-dependent AHC calculated from the Berry curvature exhibits a clear peak plateau, which remains above 1000 $\Omega^{-1}$ $cm^{-1}$ and shows an energy width of 100 meV just below $E_F$, as indicated by the green line in **Scheme 1**. This result implies that the contribution of the intrinsic mechanism to AHC will not significantly change against small-scale energy disturbances such as temperature change or *p*-doping. Thus, we can exploit this feature of the magnetic Weyl semimetal $Co_3Sn_2S_2$ and further dope a small amount of alien atoms onto the Co kagome lattices. In the case of Fe, which has one less valence electron than Co, it is possible to maintain the Fermi level within the energy range of the AHC plateau (black dashed line and black arrow in **Scheme 1**). Simultaneously, the asymmetric scattering of moving electrons due to Fe dopants can result in the extrinsic contribution (red dashed line and red arrows in **Scheme 1**) in addition to the large intrinsic component. In this scheme, a magnetic Weyl semimetal with clean topological bands and high AHE is necessary as it can provide an ideal platform to manipulate the extrinsic mechanism to simultaneously enhance AHC and AHA. Therefore, we selected $Co_3Sn_2S_2$ as the platform and substituted Co atoms with a small amount of Fe atoms, corresponding to the chemical formula of $Co_{3-x}Fe_xSn_2S_2$. Here, we note that Fe atoms will directly occupy Co sites in the kagome lattices in the system.

## 3. Results and Discussion

### 3.1 Basic behaviors of magnetism and electricity after Fe doping

A series of $Co_{3-x}Fe_xSn_2S_2$ (*x* = 0, 0.025, 0.05, 0.10, 0.15, 0.20) single-crystals were grown using Sn and Pb mixed flux.[28] The chemical compositions of the grown crystals were determined using energy dispersive X-ray spectroscopy (EDS) (See **Figure S1** and **Table S1**). **Figure 1a** shows the crystal structure of $Co_3Sn_2S_2$. The crystal (space group R-3m) contains a quasi-2D Co–Sn kagome layer sandwiched between S atoms and stacks in the ABC fashion along the *c*-axis. The magnetic moments of Co atoms are fixed on the kagome layer in the *ab*-plane and along the *c*-axis.[29] The room-temperature XRD patterns of $Co_{3-x}Fe_xSn_2S_2$ show only the (000*l*) Bragg peaks, which indicates that the exposed plane surface is the *ab*-plane, as shown in **Figure 1b**. Moreover, the peaks (000*l*) shift to lower diffraction angles with increase in the Fe content, as illustrated in inset of Figure 1b. This indicates that the lattice constant increases because the radius of the Fe atom is larger than that of Co. **Figures 1c and 1d** show the Fe-content dependencies of the Curie temperature ($T_C$) and 10-K saturation magnetization ($M_S$) obtained from $M(T)$ and $M(B)$, respectively (see **Figures S2a and b**). Parameter $T_C$ monotonously decreases from 175 K at *x* = 0 to 150 K at *x* = 0.20. Parameter $M_S$ also monotonously reduces from 0.92 $\mu_B$/f.u. at *x* = 0 to 0.72 $\mu_B$/f.u. at *x* = 0.20. Both these parameters decrease with increase in the Fe doping amount, which is mainly attributed to the weaker magnetic exchange of Fe relative to Co. **Figure 1e** shows the longitudinal resistivity $\rho_{xx}$ of $Co_{3-x}Fe_xSn_2S_2$ at zero field for I // *a*. All the curves exhibit a kink point corresponding to $T_C$, determined upon deriving the $\rho_{xx}$ curves (see details in **Figure S2c)**. Furthermore, the samples with *x* = 0.15 and 0.20 demonstrate the Kondo effect[30,31] below 50 K, which is caused by the magnetic scattering due to the presence of the doped magnetic element Fe.

### 3.2 Fe-doping dependence of AHE

**Figure 2a** shows Hall resistivity $\rho_{yx}$ as a function of magnetic field $B$ at 10 K for $B // c$ and $I // a$. The anomalous Hall resistivity $\rho_{yx}^A$ at 10 K is obtained at zero field. We plot $\rho_{yx}^A$ and $\rho_{xx}$ as functions of the Fe content in **Figure 2b**. With increase in the Fe doping amount, $\rho_{yx}^A$ increases gradually from 4 μΩ cm at $x = 0$ to 98 μΩ cm at $x = 0.20$, corresponding to an increment of two orders in amplitude. This giant $\rho_{yx}^A$ value is seldom observed in AHE materials, particularly in single-crystalline materials. Parameter $\rho_{xx}$ increases gradually from 69 μΩ cm at $x = 0$ to 320 μΩ cm at $x = 0.20$ at 10 K, which is caused by the enhancement of impurity scattering upon Fe doping. Furthermore, AHC $\sigma_{xy}^A$ can be calculated from $\rho_{xx}$ and $\rho_{yx}$ via the relation $\sigma_{xy}^A = -\sigma_{yx}^A = \rho_{yx}/(\rho_{yx}^2 + \rho_{xx}^2)$, as shown in **Figure 2c**. Parameter $\sigma_{xy}^A$ shows an initial increment and then decreases with increase in the Fe doping amount, exhibiting a maximum of 1850 $\Omega^{-1}$ cm$^{-1}$ at $x = 0.05$, which is nearly twice as large as that of undoped $Co_3Sn_2S_2$. In addition, the AHA characterized by $\sigma_{xy}^A/\sigma_{xx}$ is also an important measure to describe the transport properties of AHE materials, and it reflects the efficiency of the conversion of longitudinal current into transverse current, e.g., anomalous Hall current. The AHA at 10 K also demonstrates a similar Fe-doping dependence, as shown in **Figure 2d**, exhibiting a maximum of ~33% at $x = 0.15$ under zero field, which is considerably larger than that of any magnetic bulk material reported previously.

**Figure 2e** shows the temperature dependence of the AHC at 1 kOe (larger than the out-of-plane saturation field of 0.9 kOe)[29] from 10 to 300 K (also see **Figure S3**). As the AHC at $x = 0$ is basically dominated by the Berry curvature of the topological band structures, the AHC remains unchanged below 100 K.[17] With further increase in temperature, the increase in thermal disturbance initiates the random arrangement of magnetic moments, which reduces the strength of the Berry curvature and AHC.[17,32,33] Above $T_C$, the AHC disappears because the magnetic moment is completely disordered. For $x = 0.025, 0.05, 0.10, 0.15$, and $0.20$, the AHC decreases with increase in temperature; this result is discussed later in the paper.

The AHA, characterized by the ratio of $\sigma_{xy}^A/\sigma_{xx}$, strongly depends on the synergetic change of $\sigma_{xy}^A$ and $\sigma_{xx}$ (see Figure S4). For x = 0, the topologically protected $\sigma_{xy}^A$ is relatively robust against temperature in low temperatures. In contrast, the $\sigma_{xx}$ is sensitive to temperature owing to electron-phonon scattering and decreases rapidly with the increase of temperature. Therefore, The AHA increases gradually with the increase of temperature. However, with the further increase of temperature, the $\sigma_{xy}^A$ decreases significantly, leading to the decrease of the AHA and showing a maximum of 19% at 135 K. The AHA of x = 0.025 shows the same trend like that of x = 0, showing a maximum of 23.6% at 110 K. For the AHA of x = 0.05, because both $\sigma_{xy}^A$ and $\sigma_{xx}$ decrease at low temperatures with the increase of temperature, the AHA basically keeps at ~22% unchanged within the range of 100 K. For the AHA of x = 0.10, 0.15 and 0.20, the $\sigma_{xy}^A$ decreases significantly with the increase of temperature, while the change of the $\sigma_{xx}$ is not significant. Therefore, the AHAs of x = 0.10, 0.15 and 0.20 decrease all the time with increase of the temperature, showing a maximum ~30%, ~33% and ~31%, respectively, at 10 K. It should be noted here that the AHC and AHA are very large over a wide temperature range.

### 3.3 Separation of intrinsic and extrinsic mechanisms via application of TYJ model

Till now, we have obtained higher measured vales of AHC and AHA than those of undoped $Co_3Sn_2S_2$. To verify that the enhanced AHE in $Co_{3-x}Fe_xSn_2S_2$ is caused by the increase in the extrinsic contribution, we require an effective model for separating the intrinsic and extrinsic contributions. It is generally accepted that the total $\sigma_{xy}^A$ can be expressed as the sum of three terms: $\sigma_{xy}^A = \sigma_{int} + \sigma_{sk} + \sigma_{sj}$, where $\sigma_{int}$, $\sigma_{sk}$, and $\sigma_{sj}$ denote the intrinsic, skew, and side-jump contributions, respectively. In this regard, Tian et al. proposed the following scaling model of the AHE, $\rho_{yx}^A = a\rho_{xx0} + b\rho_{xx}^2$. This is the so-called TYJ scaling,[27] which has been used to successfully distinguish these three contributions in many systems.[34-39] In the equation, $\rho_{xx0}$ denotes the residual resistivity. In the TYJ scaling model, the first term on the right-hand side of the equation corresponds to the extrinsic contributions, which include the skew and side-jump components. The second term corresponds to the intrinsic contribution. The TYJ scaling can also be expressed in the conductivity form as follows, $\sigma_{xy}^A = -a\sigma_{xx0}^{-1}\sigma_{xx}^2 - b$, where $\sigma_{xx0} = 1/\rho_{xx0}$ denotes the residual conductivity.

**Figure 3a** shows the $\sigma_{xy}^A$ versus $\sigma_{xx}^2$ curves for $x \leq 0.10$ for the temperature range of 10–40 K (also see **Figure S5**). we note that each curve can be fitted by a linear relationship. The intercept *b* of each line on the longitudinal axis represents the intrinsic AHC contribution. In order to display the fitting accuracy clearly, the fitting result of x = 0.05 as a representative is also shown in the **Figure S6**. Next, we separated the results into the extrinsic and intrinsic components at 10 K by applying the TYJ model, where $\sigma_{xx0} \approx \sigma_{xx}(10K)$, as shown in **Figure 3b**. It can be observed that the intrinsic $\sigma_{xy}^A$ (int.) basically lies between 850 and 1000 $\Omega^{-1}$ cm$^{-1}$ with slight variations when $x = 0$, 0.025, 0.05. This result confirms that the intrinsic contribution to AHC does not change for small amounts of Fe p-doping in $Co_3Sn_2S_2$, even if the $E_F$ value shifts slightly downward. The theoretical calculations from energy bands also reveal a consistent behavior with a slight increase (see **Figure S7** for details), which further indicates the reliability of the results separated by TYJ model. Importantly, both the intrinsic AHCs from experiment and theoretical calculations are basically kept at the same level of pristine $Co_3Sn_2S_2$, which confirms our claim that the $E_F$ is still on the AHC-plateau for the low Fe-doping levels and the intrinsic AHC will not significantly change for the low p-doping.

Meanwhile, the extrinsic $\sigma_{xy}^A$ (ext.) rapidly increases from -20 $\Omega^{-1}$ $cm^{-1}$ at $x = 0$ to 914 $\Omega^{-1}$ $cm^{-1}$ at $x = 0.05$. It is clear that slight Fe doping can produce a large, positive extrinsic contribution in the magnetic Weyl semimetal $Co_3Sn_2S_2$, while the large intrinsic component remains as the band Berry curvature is not changed by the slight doping. Therefore, giant AHC is obtained for $x = 0.05$ owing to the dual contribution of the intrinsic and extrinsic mechanisms. However, although $\sigma_{xy}^A$ (ext.) continues to increase to 1071 $\Omega^{-1}$ $cm^{-1}$ at $x = 0.10$, the AHC begins to decrease because of the further increase in Fe content, which results in a significant decrease in $E_F$, which in turn corresponds to a significant decrease in the intrinsic contribution, as depicted by the intrinsic AHC line (green) in **Scheme 1**. As regards the origin of the extrinsic contribution for $x \leq 0.10$, we make a further analysis according to $\log \rho_{yx}^A = \alpha \log \rho_{xx}$, where $\alpha$ denotes an exponent of $\rho_{xx}$ (see details in **Figure S8**). The skew and side-jump mechanisms yield relations of $\rho_H \propto \rho_{xx}$ and $\rho_H \propto \rho_{xx}^2$, respectively. With increase in Fe doping, $\alpha$ gradually decreases from 2 to 0.7, which indicates that the enhanced extrinsic contribution for $x \leq 0.10$ is mainly due to the skew mechanism. In $Co_3Sn_2S_2$, the conduction electrons mainly originate from the bands of Co in the kagome layer. The small number of doped Fe atoms randomly occupy Co sites in the kagome lattice, which leads to asymmetric scattering of the conduction electrons due to the effective spin–orbit interaction between them. Meanwhile, Wang *et al*. have reported that the side-jump contribution to AHE in $Co_3Sn_2S_2$ is negligible.[33]

With further increase in Fe doping, the Kondo effect is observed for $x = 0.15$ and $0.20$, which is indicated by the upwarps of the R–T curves at low temperatures, as shown in **Figure 1e**. Currently, no effective model can be used to separate the AHC in Kondo systems. A previous work has reported that the Kondo effect negatively contributes to AHC.[40] Moreover, a further increase in Fe doping leads to a large decrease in $E_F$, which also results in a decrease in the intrinsic contribution. Thus, the decrease in AHC at $x = 0.15, 0.20$ is mainly because of the decrease in $E_F$ and the Kondo effect. In addition, the temperature dependences of the total $\sigma_{xy}^A$, $\sigma_{xy}^A$ (int.), and $\sigma_{xy}^A$ (ext.) at $x = 0.05$ from 10 to 40 K at zero field are shown in **Figure 3c**. With increase in temperature over a small range, $\sigma_{xy}^A$ (int.) remains at around 930 $\Omega^{-1}$ $cm^{-1}$, as discussed above. However, with increase in temperature, the phonon scattering increases and the longitudinal conductivity decreases, which reduces the extrinsic contribution with increase in temperature, as per ($-a\sigma_{xx0}^{-1}\sigma_{xx}^2$). Therefore, $\sigma_{xy}^A$ (ext.) and total $\sigma_{xy}^A$ decrease with increase in temperature, while $\sigma_{xy}^A$ (int.) remains constant, as illustrated in **Figure 3d**.

### 3.4 Comparison of Co$_{3-x}$Fe$_x$Sn$_2$S$_2$ with other AHE materials

In most materials, the AHC mainly originates from the topologically trivial electronic bands and impurity scattering. A typical feature of these materials is that both $\sigma_{xy}^A$ and $\sigma_{xx}$ are either large or small, and therefore, AHC and AHA are not large at the same time. Meanwhile, the magnetic Weyl semimetal Co$_3$Sn$_2$S$_2$, owing to the topological-band-structure-induced large Berry curvature and the Weyl semi-metallic character, possesses both large AHC and AHA simultaneously at zero magnetic field.[17]

**Figure 4** shows the AHA and AHC for different Fe-doping levels of Co$_{3-x}$Fe$_x$Sn$_2$S$_2$ in the temperature range from 150 to 10 K along with those of other materials measured in low magnetic fields. It is clear that the AHA and AHC values of most AHE materials are not simultaneously large, as indicated by their location in the lower left corner of the plot in **Figure 4.** On the contrary, both the AHC and AHA values of Co$_{3-x}$Fe$_x$Sn$_2$S$_2$ are the largest of all AHE materials (as indicated by the plots located in the upper right corner). Moreover, it should be noted that the AHA at *x* = 0.15 is as high as ~33%, which is the largest AHA of known materials previously reported in low magnetic fields. Such large values of AHC and AHA are derived by increasing the extrinsic contribution over and above the constant intrinsic contribution. Moreover, Co$_{3-x}$Fe$_x$Sn$_2$S$_2$ can further maintain these large AHC and AHA values over a wide temperature range. The large AHA at zero field indicates that up to one-third of the longitudinal driving current can be converted to transverse current. More importantly, the huge AHA is obtained in zero field due to the low saturation field and large magnetic coercivity, which is quite different from the case of other systems showing AHA only at high fields. All of these can be greatly beneficial for potential applications such as AHE sensors[4] or those based on spin-transfer torque.[5]

## 4. Conclusion

In summary, we proposed a scheme to simultaneously tune the AHC and AHA to large values in the magnetic Weyl semimetal Co$_{3-x}$Fe$_x$Sn$_2$S$_2$ by means of *p*-doping. Exploiting the fact that the large intrinsic AHC contribution in Co$_3$Sn$_2$S$_2$ is basically unchanged for slight *p*-doping, we considered that a small amount of Fe (with a lesser number of electrons relative to Co) replacing Co in the magnetic Weyl semimetal Co$_3$Sn$_2$S$_2$ can increase the extrinsic contribution via asymmetric impurity scattering. Thus, significantly enhanced AHC (~1850 $\Omega^{-1}$ cm$^{-1}$) and AHA (~33%) values were obtained in the magnetic Weyl semimetal Co$_{3-x}$Fe$_x$Sn$_2$S$_2$, and these values persisted over a wide temperature range. Here, we note that the AHA value (~33%) at zero field is a new record among known magnetic materials. The obtained large anomalous Hall current over a wide temperature range can be exploited for application in topological spintronics or Weyltronics. Furthermore, our study provides an effective strategy to manipulate the giant AHE via controlling the intrinsic and/or extrinsic contributions in magnetic Weyl materials.

**Single crystal growth.** The single crystals of $Co_{3-x}Fe_xSn_2S_2$ ($0 \leq x \leq 0.20$) can be grown using Sn and Pb mixed flux. Co (99.95% Alfa), Fe (99.95% Alfa), Sn (99.999% Alfa), S (99.999% Alfa) and Pb (99.999% Alfa) were mixed in an initial mixture of molar ratios (Co+Fe) : S : Sn : Pb = 12 : 8 : 35 : 45. These mixtures were placed in a $Al_2O_3$ crucible. Then, the $Al_2O_3$ crucibles are sealed in quartz tubes that were put in an electric furnace. The quartz tubes were slowly heated to 673 K over 6 h and held on for extra 6 h for preventing the loss of sulfur due to the high vapor pressure. Then, the quartz tubes were heated to 1323 K over 6 h and kept there for 6 h. The melt was cooled slowly to 973 K over 70 h. At 973 K, the ampoule was removed from the electric furnace and the flux was removed via rapid decanting and subsequent spinning in a centrifuge. The single crystals of $Co_{3-x}Fe_xSn_2S_2$ with typical sizes 2-5 mm and hexagonal shape were obtained. The crystal structure and composition of single crystals were confirmed by X-ray diffraction measurements and the energy dispersive X-ray spectroscopy EDS.

**Magnetization and electric transport measurements.** Magnetic properties were measured by superconducting quantum interference device (SQUID) magnetometer. Electric transport was measured by the physical property measurement system (PPMS).

**Supporting Information**
Supporting Information is available from the Wiley Online Library or from the author.

**Acknowledgements**
This work was supported by National Natural Science Foundation of China (Nos. 11974394 and 51722106), National Key R&D Program of China (Nos. 2019YFA0704904, 2017YFA0206303), the Strategic Priority Research Program (B) of the Chinese Academy of Sciences (CAS) (XDB33000000), Beijing Natural Science Foundation (No. Z190009), Users with Excellence Program of Hefei Science Center CAS (No. 2019HSC-UE009), and Fujian Institute of Innovation, CAS.

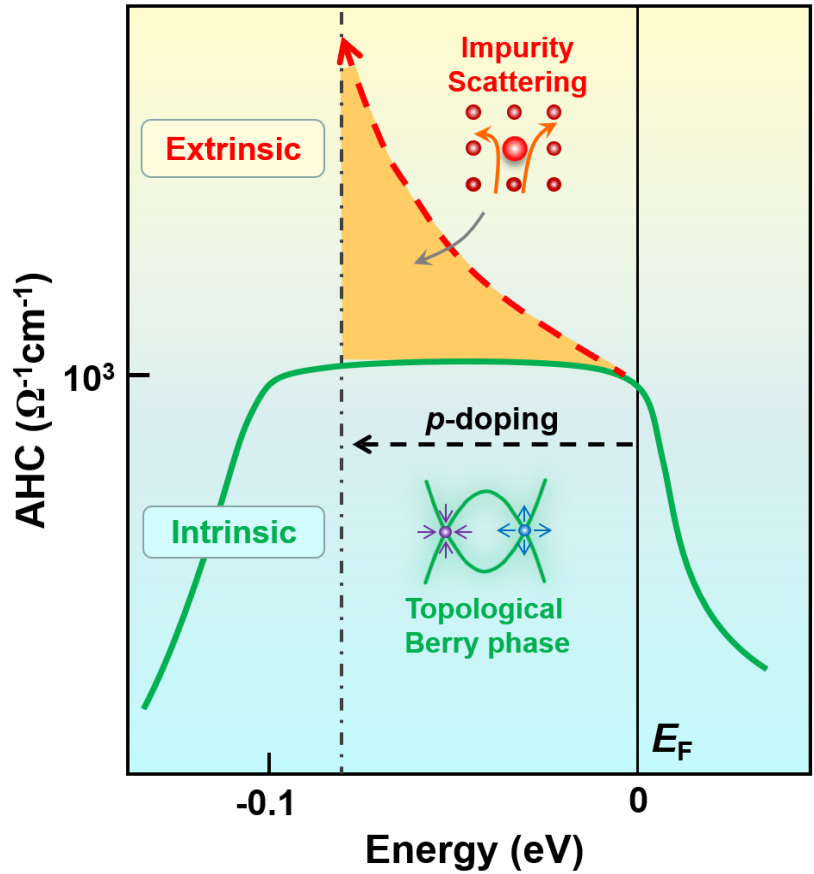

**Scheme 1. Schematic for tuning AHE via controlling intrinsic and extrinsic contributions in magnetic Weyl semimetal $Co_3Sn_2S_2$.** The green line denotes the intrinsic AHC of $Co_3Sn_2S_2$ dominated by the Berry phase from the topological Weyl bands. The red dashed line and orange area correspond to the extrinsic AHC and impurity scattering, respectively, generated by the slight p-doping of alien atoms.

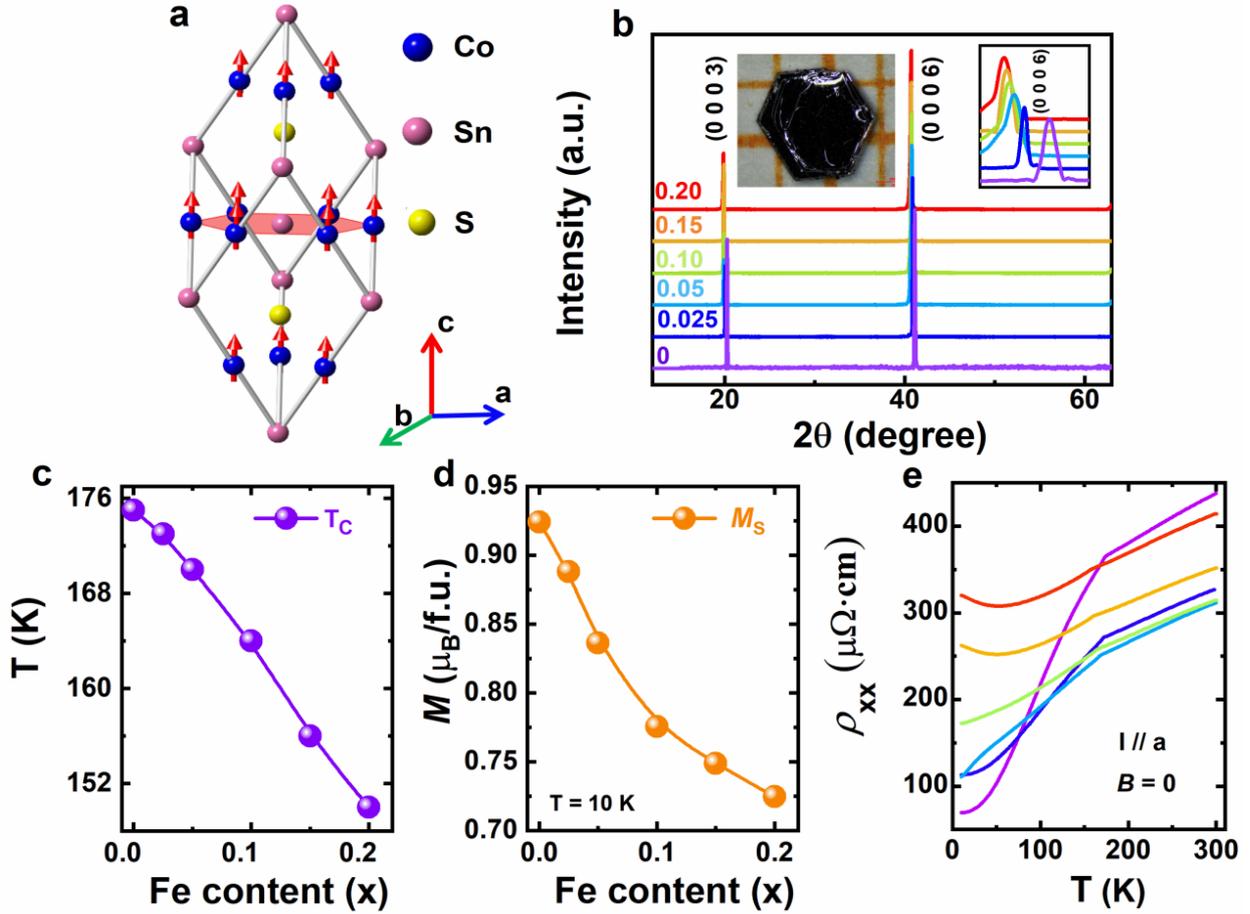

**Figure 1. Crystal structure, Curie temperature ($T_C$), saturation magnetization ($M_S$), and longitudinal resistivity ($\rho_{xx}$) of Fe-doped $Co_{3-x}Fe_xSn_2S_2$. a)** Crystal structure of $Co_3Sn_2S_2$. **b)** XRD patterns of $Co_{3-x}Fe_xSn_2S_2$ single-crystals at room temperature. The insets shows a photograph of the $Co_3Sn_2S_2$ single-crystal and a zoom-in figure on the angle scale for (0006) Bragg peaks. **c)** $T_C$ of $Co_{3-x}Fe_xSn_2S_2$. **d)** $M_S$ of $Co_{3-x}Fe_xSn_2S_2$. **e)** $\rho_{xx}$ of $Co_{3-x}Fe_xSn_2S_2$ under zero field for I // $a$.

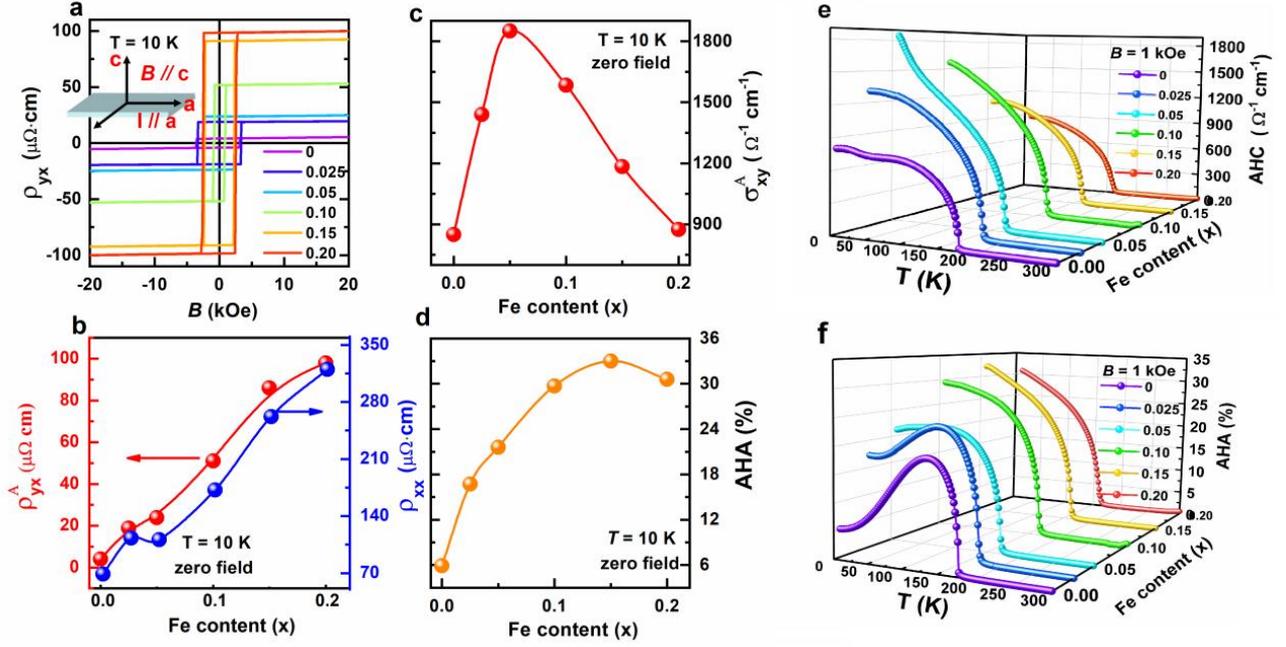

**Figure 2. AHE subsequent to Fe doping. a)** Hall resistivity $\rho_{yx}$ as a function of magnetic field $B$ at 10 K for $B \mathbin{/\mkern-5mu/} c$ and $I \mathbin{/\mkern-5mu/} a$. **b)** Fe-doping dependences of anomalous Hall resistivity $\rho_{yx}^A$ and longitudinal resistivity $\rho_{xx}$. **c)** and **d)** Fe-doping dependences of AHC $\sigma_{xy}^A$ and AHA, respectively, at 10 K and under zero field. **e)** and **f)** Temperature dependence of AHC and AHA values of Co$_{3-x}$Fe$_x$Sn$_2$S$_2$ under a field of 1 kOe.

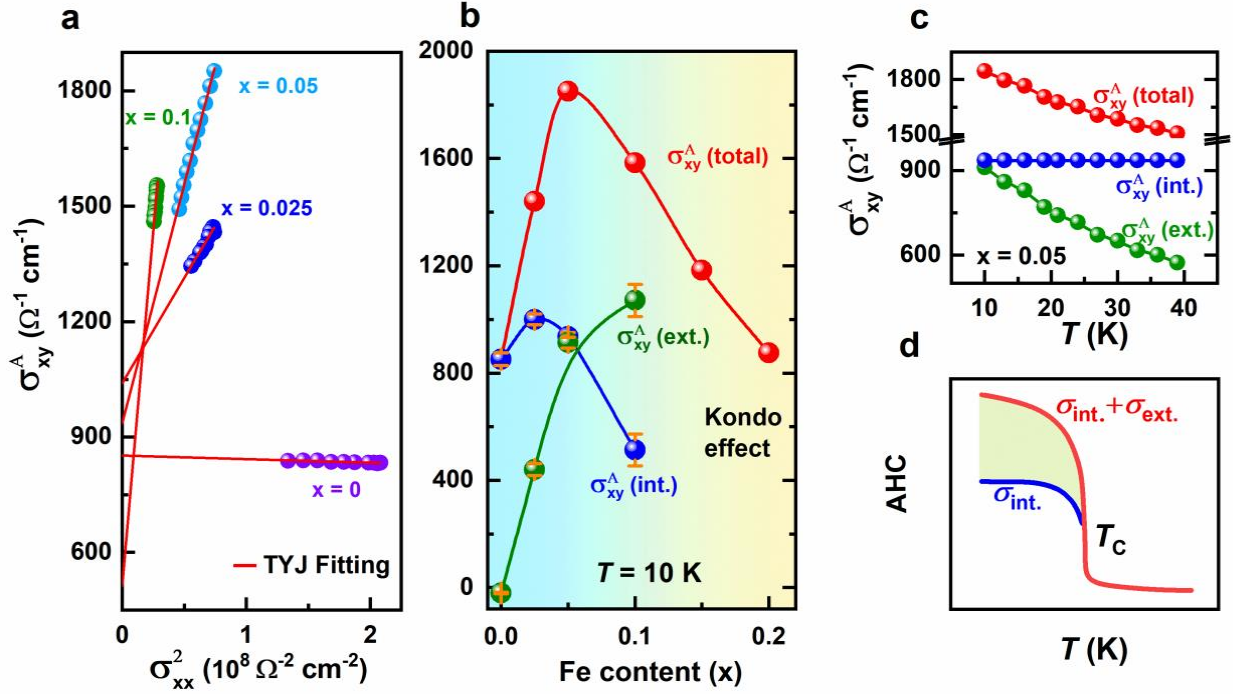

**Figure 3. Scaling of AHE. a)** Parameter $\sigma_{xy}^{A}$ as function of square of longitudinal conductivity $\sigma_{xx}^{2}$. The solid red lines indicate the fitting results with $\sigma_{xy}^{A} = -a\sigma_{xx0}^{-1}\sigma_{xx}^{2} - b$. **b)** Scaling AHC data of Co$_{3-x}$Fe$_{x}$Sn$_{2}$S$_{2}$ at 10 K. **c)** Temperature dependence of total AHC $\sigma_{xy}^{A}$, intrinsic $\sigma_{xy}^{A}$ (int.), and extrinsic $\sigma_{xy}^{A}$ (ext.) at $x = 0.05$ from 10 to 40 K at zero field. **d)** Schematic of the temperature dependence of $\sigma_{xy}^{A}$, $\sigma_{xy}^{A}$ (int.), and $\sigma_{xy}^{A}$ (ext.).

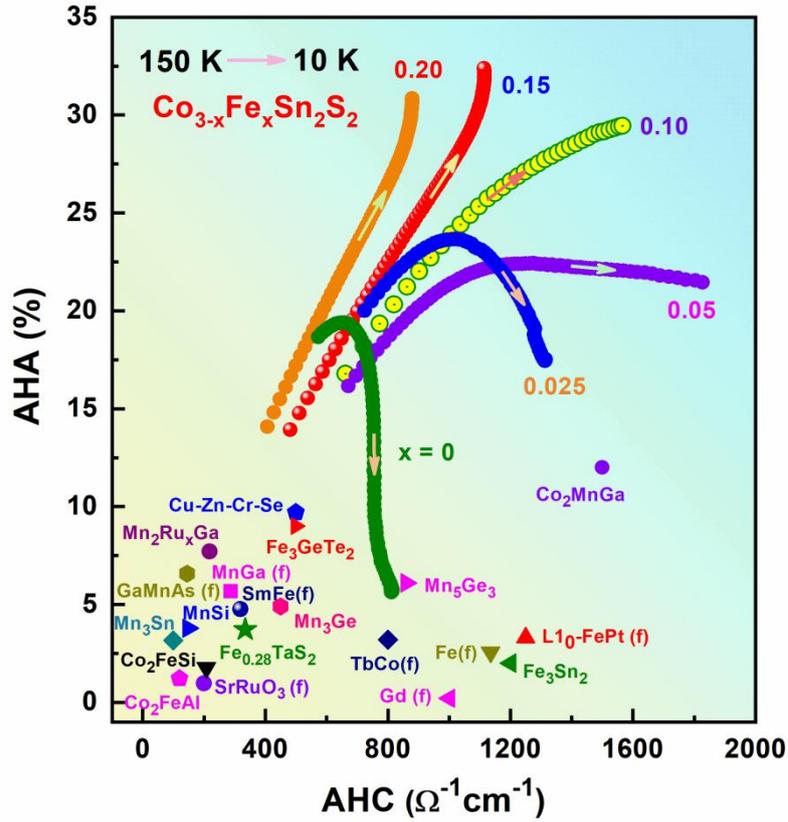

**Figure 4. Comparison of AHA and AHC values of $Co_{3-x}Fe_xSn_2S_2$ with corresponding ones of other AHE materials.** The data for different doping levels of $Co_{3-x}Fe_xSn_2S_2$ are shown from 150 to 10 K by means of arrows on the curves. '(f)' denotes thin-film materials. From the view of the potential applications, the applied external magnetic field for Hall response should be quite low, which means that the ground state of the material at zero field is FM or sometimes, AFM but with nonzero Berry curvature such as $Mn_3Sn$ or $Mn_3Ge$. All the materials are selected based on this principle.